\RequirePackage{fix-cm}
\documentclass[smallextended]{svjour3}       
\smartqed  
\usepackage{graphicx}
\usepackage{enumerate}
\usepackage{float}

\newcommand{\tabincell}[2]{\begin{tabular}{@{}#1@{}}#2\end{tabular}}

\begin{document}

\title{Three-party quantum private comparison of equality based on genuinely maximally entangled six-qubit states}
\author{Cai Zhang \and Zhiwei Sun \and Xiang Huang \and Dongyang Long}


\institute{C. Zhang \and X. Huang \and D. Long  \at
              School of Information Science and Technology,  Sun Yat-sen University, Guangzhou, Guangdong, 510006, China \\
              \email{zhangcai.sysu@gmail.com \\
              Z. Sun  \at
              College of Information Engineering,  Shenzhen University, Shenzhen, Guangdong, 518060, China
              }
}


\maketitle

\begin{abstract}
We propose a new three-party quantum private comparison protocol using genuinely maximally entangled six-qubit states. In our protocol, three participants can determine whether their private information are equal or not without an external third party who helps compute the comparison result. At the same time the participants can preserve the privacy of their inputs, respectively. Our protocol does not need any unitary operations to encode information due to the excellent properties of genuinely maximally entangled six-qubit states. Additionally, the protocol uses one-step quantum transmission and it is congenitally free from Trojan horse attacks. We have also shown that our protocol is secure against outside and participant attacks in this paper.
\keywords{Three-party quantum private comparison\and Genuinely maximally entangled six-qubit state \and Trojan horse attack}
\end{abstract}

\section{Introduction}
\label{intro}
The first quantum key distribution (QKD) protocol was proposed by Bennett and Brassard\cite{BB84} in 1984. After that, an increasing number of quantum cryptographic protocols such as quantum secret sharing (QSS) \cite{HBB99,KKI99,KB02,ZL07,JWGQG12,S12,long2012quantum}, quantum secure direct communication (QSDC) \cite{BF02,DLL03,GYW05,LWGZ08,LCJ12,sun2012quantum,sun2012quantum1}, quantum key agreement (QKA) \cite{zhou2004quantum,chong2010quantum,liu2013multiparty,sun2013improvements}, quantum summation \cite{H02,HN03,DCWZ07,CXYW10,zhang2014high}  and quantum private comparison (QPC) \cite{yang2009efficient,chen2010efficient,tseng2012new} have been presented. Quantum private comparison, as a subfield of quantum cryptography, has attracted more and more researchers. The aim of QPCs is to compare the participants' private information without publicly revealing their respective private information.\par
Since Yao \cite{yao1982protocols} presented a protocol for the millionaires' problem in which the participants try to determine which one is richer without revealing their actual wealth, the protocols of private comparison have widely been investigated. Boudot et al. \cite{boudot2001fair} proposed a protocol to decide whether two millionaires are equally rich or not. But Lo \cite{Lo97} showed that the equality function cannot be securely evaluated in a two-party scenario. Thus, some additional assumptions such as a semi-honest third party are required to achieve the goal of private comparison.\par
Yang et al. \cite{yang2009efficient} proposed the first quantum private comparison protocol in which the entanglement of Einstein-Podolsky-Rosen (EPR) pairs and the one-way hash function are employed. Chen et al. \cite{chen2010efficient} presented an efficient protocol of equality using triplet states. Ref.\cite{tseng2012new} gave a more efficient quantum private comparison of equality protocol without the entanglement of EPR pairs. Liu et al. \cite{liu2011efficient,liu2012protocol,liu2012new,liu2012quantum} presented QPC protocols employing the triplet W states, $\chi$-type genuine four-particle entangled states and the Bell states. Huang et al. \cite{huang2013robust} designed a quantum private comparison of equality protocol with collective detection-noise channels. Liu et al. \cite{liu2013efficient} employed single photons and collective detection to devise an efficient quantum private comparison protocol. Zhang et al. \cite{zhang2013quantum} proposed a quantum private comparison protocol based on an quantum search algorithm. Li et al. \cite{li2014efficient} presented an efficient protocol for equal information comparison based on four-particle entangled W state and Bell entangled states swapping. All the above protocol works for two-party who wish to compare their private information. Recently, some multi-user quantum private comparison protocols were presented. Chang et al. \cite{chang2013multi} gave a multi-user private comparison protocol using GHZ class states. Liu et al. \cite{liu2013multi} presented a multi-party quantum private comparison protocol using $d$-dimensional basis states without entanglement swapping. \par
All the above protocols include a third party who helps participants compute comparison results. Lin et al. \cite{lin2014quantum} used EPR pairs and a one-way hash function to design a quantum private comparison of equality protocol without a third party that works for the two-party scenario. If the number of participants exceeds 2, can one find a protocol for quantum private comparison without an external third party who helps compute the comparison result? We find such a protocol for three-party quantum private comparison. The protocol is based on genuinely maximally entangled six-qubit states (we name it BPB state). In our protocol, three participants can determine whether their private information are equal without the external third party, and meanwhile preserve the privacy of their inputs, respectively. Any unitary operations are not required to encode information due to the excellent properties of genuinely maximally entangled six-qubit states. Furthermore, the protocol employs one-step quantum transmission, hence it will not suffer Trojan horse attacks. Our protocol is also proven to be secure against various attacks including outside and participant attacks.\par
The rest of this paper is organized as follows. In Section 2, we analyze the structure of the genuinely maximally BPB state and show the excellent properties which are useful for designing our protocol. In Section 3, we propose a protocol of quantum private comparison based on genuinely maximally BPB states. In Section 4, we analyze the correctness and the security of the presented protocol. Finally, we make a conclusion in Section 5.

\section{The genuinely maximally entangled six-qubit state}

Quantum entanglement, as a physical resource, plays a key role in many applications such as quantum teleportation \cite{bennett1993teleporting}, quantum dense coding \cite{bennett1992communication}, quantum key distribution \cite{ekert1991quantum}, quantum secret sharing \cite{cleve1999share} and quantum secure direct communication \cite{sun2012quantum,sun2012quantum1}. \par
By using a numeric searching program, Borras et al. \cite{borras2007multiqubit} found the genuinely maximally BPB state, which is
\begin{eqnarray}
\frac{1}{\sqrt{32}} [ (|000 000\rangle &+& |111 111\rangle + |000011\rangle + |111100\rangle \nonumber\\
&+& |000101\rangle + |111010\rangle + |000110\rangle + |111001\rangle \nonumber\\
&+& |001001\rangle + |110110\rangle + |001111\rangle + |110000\rangle \nonumber\\
&+& |010001\rangle + |101110\rangle + |010010\rangle + |101101\rangle \nonumber\\
&+& |011000\rangle + |100111\rangle + |011101\rangle + |100010\rangle) \nonumber\\
&-&( |010100\rangle + |101011\rangle + |010111\rangle + |101000\rangle \nonumber\\
&+& |011011\rangle + |100100\rangle + |001010\rangle + |110101\rangle \nonumber\\
&+& |001100\rangle + |110011\rangle + |011110\rangle + |100001\rangle)]_{123456}.
\end{eqnarray}
We denote this six-qubit state by $\Psi_{6qb}$. From the above formula, we can see that $\Psi_{6qb}$ includes 32 terms, each of which has even $|0\rangle$ and equal coefficient.\par

To show the engtangled property of $\Psi_{6qb}$, we can rewrite it as

\begin{eqnarray}\label{first}
\Psi_{6qb} =
\frac{1}{\sqrt{8}} [|000\rangle|\gamma_{1}^{1}\rangle &+& |001\rangle|\gamma_{2}^{1}\rangle + |010\rangle|\gamma_{3}^{1}\rangle + |011\rangle|\gamma_{4}^{1}\rangle \nonumber\\
&-& |100\rangle|\gamma_{5}^{1}\rangle - |101\rangle|\gamma_{6}^{1}\rangle + |110\rangle|\gamma_{7}^{1}\rangle \nonumber \\
&+& |111\rangle|\gamma_{8}^{1}\rangle
 ]_{123456},
\end{eqnarray}
where $\{|\gamma_{j}^{1}\rangle|j=1,2,\ldots,8\}$ forms an orthogonal basis on Hilbert space $C_{2} \otimes C_{2} \otimes C_{2}$ such that
\begin{eqnarray}
|\gamma_{1}^{1}\rangle = \frac{1}{\sqrt{2}}(|0\rangle|\Phi^{+}\rangle + |1\rangle|\Psi^{+}\rangle), \
|\gamma_{2}^{1}\rangle = \frac{1}{\sqrt{2}}(|0\rangle|\Psi^{-}\rangle - |1\rangle|\Phi^{-}\rangle), \nonumber \\
|\gamma_{3}^{1}\rangle = \frac{1}{\sqrt{2}}(|0\rangle|\Psi^{+}\rangle - |1\rangle|\Phi^{+}\rangle), \
|\gamma_{4}^{1}\rangle = \frac{1}{\sqrt{2}}(|0\rangle|\Phi^{-}\rangle + |1\rangle|\Psi^{-}\rangle), \nonumber \\
|\gamma_{5}^{1}\rangle = \frac{1}{\sqrt{2}}(|0\rangle|\Psi^{-}\rangle + |1\rangle|\Phi^{-}\rangle), \
|\gamma_{6}^{1}\rangle = \frac{1}{\sqrt{2}}(|0\rangle|\Phi^{+}\rangle - |1\rangle|\Psi^{+}\rangle), \nonumber \\
|\gamma_{7}^{1}\rangle = \frac{1}{\sqrt{2}}(|0\rangle|\Phi^{-}\rangle - |1\rangle|\Psi^{-}\rangle), \
|\gamma_{8}^{1}\rangle = \frac{1}{\sqrt{2}}(|0\rangle|\Psi^{+}\rangle + |1\rangle|\Phi^{+}\rangle),
\end{eqnarray}
and $|\Phi^{\pm}\rangle$ and $|\Psi^{\pm}\rangle$ are Bell states in the form of
\begin{eqnarray}\label{BellStates}
|\Phi^{+}\rangle = \frac{1}{\sqrt{2}}(|00\rangle + |11\rangle) = \frac{1}{\sqrt{2}}(|++\rangle + |--\rangle), \label{firstBell} \\
|\Phi^{-}\rangle = \frac{1}{\sqrt{2}}(|00\rangle - |11\rangle) = \frac{1}{\sqrt{2}}(|+-\rangle + |-+\rangle),\\
|\Psi^{+}\rangle = \frac{1}{\sqrt{2}}(|01\rangle + |10\rangle) = \frac{1}{\sqrt{2}}(|++\rangle - |--\rangle),\\
|\Psi^{-}\rangle = \frac{1}{\sqrt{2}}(|01\rangle - |10\rangle) = \frac{1}{\sqrt{2}}(|-+\rangle - |+-\rangle). \label{lastBell}
\end{eqnarray}
Here, $|+\rangle=\frac{1}{\sqrt{2}}(|0\rangle+|1\rangle)$ and $|-\rangle=\frac{1}{\sqrt{2}}(|0\rangle-|1\rangle)$ are the eigenvectors of the Pauli operator $\sigma_{X}$.\par

We can also rewrite $\Psi_{6qb}$ as
\begin{eqnarray}\label{second}
\Psi_{6qb}=
\frac{1}{\sqrt{8}} [|+++\rangle|\gamma_{1}^{2}\rangle &+& |++-\rangle|\gamma_{2}^{2}\rangle + |+-+\rangle|\gamma_{3}^{2}\rangle + |+--\rangle|\gamma_{4}^{2}\rangle \nonumber\\
&+& |-++\rangle|\gamma_{5}^{2}\rangle + |-+-\rangle|\gamma_{6}^{2}\rangle + |--+\rangle|\gamma_{7}^{2}\rangle \nonumber\\
&+& |---\rangle|\gamma_{8}^{2}\rangle
 ]_{123456},
\end{eqnarray}
where $\{|\gamma_{j}^{2}\rangle|j=1,2,\ldots,8\}$ also forms an orthogonal basis on Hilber space $C_{2} \otimes C_{2} \otimes C_{2}$ such that
\begin{eqnarray}
|\gamma_{1}^{2}\rangle = \frac{1}{\sqrt{2}}(|+\rangle|\Psi^{+}\rangle + |-\rangle|\Phi^{-}\rangle), \
|\gamma_{2}^{2}\rangle = \frac{1}{\sqrt{2}}(-|+\rangle|\Psi^{-}\rangle + |-\rangle|\Phi^{+}\rangle), \nonumber \\
|\gamma_{3}^{2}\rangle = \frac{1}{\sqrt{2}}(-|+\rangle|\Phi^{-}\rangle - |-\rangle|\Psi^{+}\rangle), \
|\gamma_{4}^{2}\rangle = \frac{1}{\sqrt{2}}(|+\rangle|\Phi^{+}\rangle - |-\rangle|\Psi^{-}\rangle), \nonumber \\
|\gamma_{5}^{2}\rangle = \frac{1}{\sqrt{2}}(|+\rangle|\Psi^{-}\rangle + |-\rangle|\Phi^{+}\rangle), \
|\gamma_{6}^{2}\rangle = \frac{1}{\sqrt{2}}(|+\rangle|\Psi^{+}\rangle - |-\rangle|\Phi^{-}\rangle), \nonumber \\
|\gamma_{7}^{2}\rangle = \frac{1}{\sqrt{2}}(|+\rangle|\Phi^{+}\rangle + |-\rangle|\Psi^{-}\rangle), \
|\gamma_{8}^{2}\rangle = \frac{1}{\sqrt{2}}(|+\rangle|\Phi^{-}\rangle - |-\rangle|\Psi^{+}\rangle),
\end{eqnarray}
where $|\Phi^{\pm}\rangle$ and $|\Psi^{\pm}\rangle$ are Bell states as Eqs.(\ref{firstBell}-\ref{lastBell}) and  $|+\rangle=\frac{1}{\sqrt{2}}(|0\rangle+|1\rangle)$ and $|-\rangle=\frac{1}{\sqrt{2}}(|0\rangle-|1\rangle)$ are the eigenvectors of the Pauli operator $\sigma_{X}$.\par
We will use Eq. (\ref{first}) and Eq. (\ref{second}) in our protocol to check whether the participant who distributes the state $\Psi_{6qb}$ is honest or not. \par
Let us further investigate the properties of the state $\Psi_{6qb}$.
    \begin{eqnarray}
    \Psi_{6qb}&=&\frac{1}{2}(|\Phi^{+}\rangle_{12}|\Phi^{+}\rangle_{36}|\Phi^{+}\rangle_{45}
    +|\Phi^{-}\rangle_{12}|\Psi^{-}\rangle_{36}|\Psi^{+}\rangle_{45} \nonumber\\
    &+&|\Psi^{-}\rangle_{12}|\Psi^{+}\rangle_{36}|\Phi^{-}\rangle_{45}
    +|\Psi^{+}\rangle_{12}|\Phi^{-}\rangle_{36}|\Psi^{-}\rangle_{45}) \label{twoSplit1}\\
    &=&\frac{1}{2}(-|\Phi^{-}\rangle_{13}|\Phi^{-}\rangle_{24}|\Phi^{+}\rangle_{56}
    +|\Phi^{+}\rangle_{13}|\Psi^{+}\rangle_{24}|\Psi^{+}\rangle_{56} \nonumber\\
    &-&|\Psi^{+}\rangle_{13}|\Psi^{-}\rangle_{24}|\Phi^{-}\rangle_{56}
    -|\Psi^{-}\rangle_{13}|\Phi^{+}\rangle_{24}|\Psi^{-}\rangle_{56}) \\
    &=&\frac{1}{2}(|\Phi^{-}\rangle_{14}|\Phi^{+}\rangle_{26}|\Phi^{-}\rangle_{35}
    +|\Phi^{+}\rangle_{14}|\Psi^{+}\rangle_{26}|\Psi^{+}\rangle_{35} \nonumber\\
    &+&|\Psi^{-}\rangle_{14}|\Psi^{-}\rangle_{26}|\Phi^{+}\rangle_{35}
    +|\Psi^{+}\rangle_{14}|\Phi^{-}\rangle_{26}|\Psi^{-}\rangle_{35}) \\
    &=&\frac{1}{2}(|\Phi^{+}\rangle_{15}|\Phi^{+}\rangle_{23}|\Phi^{+}\rangle_{46}
    +|\Phi^{-}\rangle_{15}|\Psi^{+}\rangle_{23}|\Psi^{-}\rangle_{46} \nonumber\\
    &+&|\Psi^{+}\rangle_{15}|\Psi^{-}\rangle_{23}|\Phi^{-}\rangle_{46}
    +|\Psi^{-}\rangle_{15}|\Phi^{-}\rangle_{23}|\Psi^{+}\rangle_{46}) \\
    &=&\frac{1}{2}(|\Phi^{-}\rangle_{16}|\Phi^{+}\rangle_{25}|\Phi^{-}\rangle_{34}
    +|\Phi^{+}\rangle_{16}|\Psi^{-}\rangle_{25}|\Psi^{-}\rangle_{34} \nonumber\\
    &+&|\Psi^{+}\rangle_{16}|\Psi^{+}\rangle_{25}|\Phi^{+}\rangle_{34}
    +|\Psi^{-}\rangle_{16}|\Phi^{-}\rangle_{25}|\Psi^{+}\rangle_{34}). \label{twoSplit5}
    \end{eqnarray}
    From the above Eqs.(\ref{twoSplit1}-\ref{twoSplit5}), it is obvious to see that the other four qubits will collapse to the tensor product of two pairs of EPR when any two qubits of $\Psi_{6qb}$ are measured with the Bell Basis $\{|\Phi^{+}\rangle,|\Phi^{-}\rangle,|\Psi^{+}\rangle,|\Psi^{-}\rangle\}$. However, these two-split forms of $\Psi_{6qb}$ are not suitable for our task. We should rewrite $\Psi_{6qb}$ as
    \begin{eqnarray}\label{beUsed}
        \Psi_{6qb}=\frac{1}{4}[
            |\Phi^{+}\rangle_{12}(&|\Phi^{+}\rangle_{34}&|\Phi^{+}\rangle_{65}
                                +|\Phi^{-}\rangle_{34}|\Phi^{-}\rangle_{65} \nonumber\\
                                +&|\Psi^{+}\rangle_{34}&|\Psi^{+}\rangle_{65}
                                +|\Psi^{-}\rangle_{34}|\Psi^{-}\rangle_{65}) \nonumber\\
            +|\Phi^{-}\rangle_{12}(-&|\Phi^{+}\rangle_{34}&|\Phi^{-}\rangle_{65}
                                +|\Phi^{-}\rangle_{34}|\Phi^{+}\rangle_{65} \nonumber\\
                                -&|\Psi^{+}\rangle_{34}&|\Psi^{-}\rangle_{65}
                                +|\Psi^{-}\rangle_{34}|\Psi^{+}\rangle_{65}) \nonumber\\
            +|\Psi^{+}\rangle_{12}(&|\Phi^{+}\rangle_{34}&|\Psi^{+}\rangle_{65}
                                +|\Phi^{-}\rangle_{34}|\Psi^{-}\rangle_{65} \nonumber\\
                                -&|\Psi^{+}\rangle_{34}&|\Phi^{+}\rangle_{65}
                                -|\Psi^{-}\rangle_{34}|\Phi^{-}\rangle_{65}) \nonumber\\
            +|\Psi^{-}\rangle_{12}(-&|\Phi^{+}\rangle_{34}&|\Psi^{-}\rangle_{65}
                                +|\Phi^{-}\rangle_{34}|\Psi^{+}\rangle_{65} \nonumber\\
                                +&|\Psi^{+}\rangle_{34}&|\Phi^{-}\rangle_{65}
                                -|\Psi^{-}\rangle_{34}|\Phi^{+}\rangle_{65})].
    \end{eqnarray}

Let us agree on the following encoding:
\begin{eqnarray}
  &|\Phi^{+}\rangle& \rightarrow 00, \quad |\Phi^{-}\rangle \rightarrow 01, \nonumber \\
  &|\Psi^{+}\rangle& \rightarrow 10, \quad |\Psi^{-}\rangle \rightarrow 11.  \label{bEncoding}
\end{eqnarray}
We denote the encoding of $x$ as $Encod(x)$ where $x \in \{ |\Phi^{+}\rangle, |\Phi^{-}\rangle, |\Psi^{+}\rangle,  |\Psi^{-}\rangle \}$. For example, $Encod(|\Phi^{-}\rangle)=01$. We can let $Encod(x)=Encod(-x)$ because the measurement outcome of $-x$ will be $x$ with certainty if it is measured with Bell basis. Actually, we can say that $x$ and $-x$ are the same up to a global phase factor $-1$. \par
After the above encoding Eq. (\ref{bEncoding}) , the Eq. (\ref{beUsed}) tells us that if we measure particles (1, 2) , particles (3, 4) and particles (6, 5) with Bell basis, respectively, then the responding measurement outcomes $R_{12}$, $R_{34}$ and $R_{65}$ satisfy the following equation:
\begin{equation}\label{zero}
  Encod(R_{12}) \oplus Encod(R_{34}) \oplus Encod(R_{65})=00.
\end{equation}
The Eqs.(\ref{beUsed}-\ref{zero}) allow us to design a three-party quantum comparison protocol in which three participants can determine whether their private information are equal or not without the help of an external third party and keep their inputs secret, respectively.

\section{The three-party quantum private comparison protocol}
In our protocol, we assume that the classical and quantum channels are authenticated. Suppose that three participants $P_{1}$, $P_{2}$ and $P_{3}$ have private information (secret bit strings) $M_{1}$, $M_{2}$ and $M_{3}$, respectively. They wish to determine whether  $M_{1}=M_{2}=M_{3}$ or not and preserve the privacy of their information, respectively. The length of secret bit string is $L$. We assume that the participant $P_{1}$ prepares the genuinely maximally BPB states $\Psi_{6qb}$, then the process of  the multi-party quantum private comparison protocol can be described as follows.
    \begin{enumerate}[{(S1)}]
      \item $P_{1}$ first prepares $(\lceil \frac{L}{2} \rceil + \delta)$ ($\lceil \quad \rceil$ denotes the ceiling function) genuinely maximally BPB states $\Psi_{6qb}$. Then he picks up the particles (3, 4) (particles (6, 5)) from each $\Psi_{6qb}$ to form an ordered sequence $S_{34}$ ($S_{65}$). After that $P_{1}$ prepares $d$ decoy particles, each of which is in one of the quantum states $\{ |0\rangle, |1\rangle, |+\rangle, |-\rangle \}$. He then randomly inserts the $d$ decoy particles into the sequence $S_{34}$ ($S_{65}$) to form a new sequence $S_{34}^{*}$ ($S_{65}^{*}$). Note that anyone does not know the initial states and positions of the $d$ decoy particles except $P_{1}$. At last, $P_{1}$ transmits $S_{34}^{*}$ ($S_{65}^{*}$) to participant $P_{2}$ ($P_{3}$) and keeps the ordered sequence $S_{12}$ of  particles (1, 2) from each $\Psi_{6qb}$ in his lab.
      \item \label{outCheck}Confirming that participant $P_{2}$ ($P_{3}$) has received all the particles $S_{34}^{*}$ ($S_{65}^{*}$) sent by $P_{1}$. $P_{1}$ announces the positions and the bases of the decoy particles to $P_{2}$ ($P_{3}$). In the following, participant $P_{2}$ ($P_{3}$) measures the decoy particles with one of the two bases $\{|0\rangle, |1\rangle\}$ and $\{|+\rangle, |-\rangle\}$ according to $P_{1}$'s announced information. And then $P_{2}$ ($P_{3}$) publishes his measurement outcomes. Later, $P_{1}$ can determine the error rate according to the $d$ decoy particles' initial states. If the error rate exceeds the threshold, then this protocol will be aborted and repeat the step ($S1$). Otherwise, the protocol will go to the next step.
      \item \label{step:colCheck}$P_{2}$ and $P_{3}$ collaborate to check whether $P_{1}$ distributes the intended particles to them. Namely, $P_{2}$ ($P_{3}$) should receive the ordered sequence $S_{34}$ ($S_{65}$) of particles (3, 4) (particles (6, 5)). First, $P_{2}$ ($P_{3}$) removes the decoy particles from $S_{34}^{*}$ ($S_{65}^{*}$) to get $S_{34}$ ($S_{65}$). They then randomly choose $\delta$ genuinely maximally BPB states $\Psi_{6qb}$ (we call them sample states) for checking and tell $P_{1}$ the positions of the sample states. After that they ask $P_{1}$ to measure the particles (1, 2) in each sample state with one of the two bases $\{|0\rangle, |1\rangle\}$ and $\{|+\rangle, |-\rangle\}$ randomly. If $P_{1}$ measures the particles (1, 2) with the basis $\{|0\rangle, |1\rangle\}$ ($\{|+\rangle, |-\rangle\}$), then $P_{2}$ measures the particle 3 with the basis $\{|0\rangle, |1\rangle\}$ ($\{|+\rangle, |-\rangle\}$), and $P_{2}$ and $P_{3}$ measure the particles (4, 5, 6) with the basis $\{|\gamma_{j}^{1}\rangle|j=1,2,\ldots,8\}$ ($\{|\gamma_{j}^{2}\rangle|j=1,2,\ldots,8\}$). Finally, $P_{2}$ and $P_{3}$ can determine the error rate of the correlation of their outcomes according to Eq. (\ref{first}) and Eq. (\ref{second}). If the error rate exceeds the threshold, then this protocol will be aborted and repeat the step ($S1$). Otherwise, the protocol will go to the next step.
      \item \label{step:key} By removing the particles of the sample states, $P_{1}$ ($P_{2}$, $P_{3}$) measures particles (1, 2) (particles (3, 4) , particles (6, 5)) of the $ith$ $\Psi_{6qb}$ ($i=1, 2, \ldots, \lceil \frac{L}{2} \rceil$) with the Bell basis. According to their measurement outcomes and the encoding arrangement Eq. (\ref{bEncoding}), $P_{1}$ ($P_{2}$, $P_{3}$) will get the key $K_{1}$ ($K_{2}$, $K_{3}$) that will be kept secret.  For example, the possible measurement outcomes of the $ith$ $\Psi_{6qb}$ may be $R_{12}^{i}=|\Psi^{-}\rangle$ ($R_{34}^{i}=|\Psi^{+}\rangle$, $R_{65}^{i}=|\Phi^{-}\rangle$), thus the $ith$ two bits of $K_{1}$ ($K_{2}$, $K_{3}$) is $11$ ($10$, $01$). After that $P_{1}$ ($P_{2}$, $P_{3}$) computes $C_{1}=M_{1}\oplus K_{1}$ ($C_{2}=M_{2}\oplus K_{2}$, $C_{3}=M_{3}\oplus K_{3}$) (Here, $\oplus$ denotes the addition module 2.).
      \item \label{step:last}$P_{2}$ ($P_{3}$) sends $C_{2}$ ($C_{3}$) to $P_{1}$, $P_{1}$ then can determine whether $M_{2}=M_{3}$ or not. If $M_{2}\neq M_{3}$, $P_{1}$ announces the result and the protocol finishes. Otherwise, $P_{1}$ randomly computes $C_{13}=C_{1}\oplus C_{3}$ or $C_{12}=C_{1}\oplus C_{2}$. If $P_{1}$ computes $C_{13}=C_{1}\oplus C_{3}$ ($C_{12}=C_{1}\oplus C_{2}$), he then sends $C_{13}$ ($C_{12}$) to $P_{2}$ ($P_{3}$). Subsequently $P_{2}$ ($P_{3}$) can determine whether $M_{1}=M_{3}$ ($M_{1}=M_{2}$) depending on the key $K_{2}$ ($K_{3}$) and $C_{13}$ ($C_{12}$). Finally, they can determine whether $M_{1}=M_{2}=M_{3}$ or not and preserve the privacy of their information, respectively.
    \end{enumerate}

\section{Analysis of the presented protocol}
In this section, we will analyze the correctness and the security of our protocol.
    \subsection{Correctness}
    According to the Eq. (\ref{zero}), we can find in the step (S\ref{step:key}) that $K_{1}\oplus K_{2}\oplus K_{3}=0$. When $P_{1}$ receives $C_{2}$ and $C_{3}$ from $P_{2}$ and $P_{3}$, respectively, he could compute
    \begin{eqnarray}\label{equ:P1forP23}
    K_{1}\oplus C_{2}\oplus C_{3}&=&K_{1} \oplus K_{2}\oplus M_{2} \oplus K_{3}\oplus M_{3}\nonumber \\&=& K_{1} \oplus K_{2}\oplus K_{3} \oplus M_{2}\oplus M_{3}\nonumber \\
    &=& M_{2} \oplus M_{3}.
    \end{eqnarray}
    Later, he can determine whether $M_{2} \oplus M_{3}=0$ depending on the Eq. (\ref{equ:P1forP23}). If he finds that $M_{2} \oplus M_{3}\neq 0$, he can simply announce that $M_{1}=M_{2}=M_{3}$ is false and the protocol finishes. Otherwise, the protocol will continue and $P_{2}$ ($P_{3}$) can also determine whether $M_{1}=M_{3}$ ($M_{1}=M_{2}$) or not in the step (S\ref{step:last}) using the similar method as that of $P_{1}$. Finally, the protocol could correctly determine $M_{1}=M_{2}=M_{3}$ or not.

    \subsection{Security}
    Compared with the quantum cryptography protocols such as quantum key distribution (QKD)\cite{BB84,LC99,ShorPreskill2000,He2011,FFBLSTW12}, quantum secret sharing (QSS)\cite{HBB99,KKI99,KB02,ZL07,JWGQG12,S12,long2012quantum} and quantum secure direct communication (QSDC)\cite{BF02,DLL03,GYW05,LWGZ08,LCJ12,sun2012quantum,sun2012quantum1}, the security analysis of multi-party quantum private comparison protocol is more complicated. Because the attacks from all participants have to be considered in multi-party quantum private comparison protocols. Outside eavesdroppers have the desire to get the participants' private inputs. In addition, some participants may do their utmost to derive other participants' private secret information. Therefore, multi-party quantum private comparison protocols must be secure against outside and participant attacks.\par
        \subsubsection{Outside attacks}
        Similar to the detection method for outside eavesdropping used in the BB84 QKD protocol \cite{BB84}, we employ the decoy particles to prevent the eavesdropping. It has been proven to be unconditionally secure by Ref. \cite{ShorPreskill2000}. Any outside eavesdropping will be detected in the step (S\ref{outCheck}), thus outside Eve's all kinds of attacks, such as the intercept-resend attack, the measurement-resend attack, the entanglement-measurement attack, are useless in our protocol. We take the intercept-resend attack as an example here: suppose that the initial decoy particle state is $|0\rangle$, and Eve randomly measures it with one of the two bases $\{|0\rangle, |1\rangle\}$ and $\{|+\rangle, |-\rangle\}$, and then she sends the fake particle prepared by herself according to the measurement outcomes to $P_{2}$ ($P_{3}$). Obviously, the probability of being detected during the step (S\ref{outCheck}) is $\frac{1}{4}$. When we use $d$ decoy particles for eavesdropping detection, the probability of being detected will be $ 1 - (\frac{3}{4})^{d} $. We can see that if $d$ is large enough, the probability of being detected will approach to 1. Therefore, Eve will be detected in the step (S\ref{outCheck}). \par
        On the other hand, Eve may get the ciphertexts $C_{2}$, $C_{3}$, $C_{13}$ and $C_{12}$ that are one-time pad ciphertexts in the protocol. However, she cannot get the keys $K_{1}$, $K_{2}$ or $K_{3}$ that are kept secret by the participants. Thus, she fails to derive the participants' private inputs.\par
         Trojan horse attack \cite{DLZZ05,GFKZR06,LDZ06}, such as the delay-photon Trojan horse attack and the invisible photon eavesdropping (IPE) Trojan horse attack, exists in two-way quantum communication protocols. Our protocol is congenitally free from these attacks because the presented protocol employs one-step quantum transmission.

        \subsubsection{Participant attack: one of the participants wants to steal others' inputs}

        First, we can assume that $P_{2}$ wishes to steal the input of $P_{1}$ ($P_{3}$) because the role of the participant $P_{3}$ is the same as the participant $P_{2}$. \par

        \begin{table}[htb]

        \caption{The relations of the participants' keys and their measurement outcomes.}
        \label{tab:1}

        {\begin{tabular}{llll}
        \hline\noalign{\smallskip}
         & $P_{1}$ &  $P_{2}$ & $P_{3}$  \\
        \noalign{\smallskip}\hline\noalign{\smallskip}

        \tabincell{l}{Possible measurement\\ outcomes} & \tabincell{l}{ $|\Phi^{+}\rangle_{12}$ ($|\Phi^{-}\rangle_{12}$)} & \tabincell{l}{ $|\Phi^{+}\rangle_{34}$ ($|\Phi^{+}\rangle_{34}$)} & \tabincell{l}{ $|\Phi^{+}\rangle_{65}$ ($|\Phi^{-}\rangle_{65}$)}\\
        Corresponding keys &  $00$ ($01$) & $00$ ($00$) & $00$ ($01$)\\ \\

        \tabincell{l}{Possible measurement\\ outcomes} & \tabincell{l}{ $|\Phi^{+}\rangle_{12}$ ($|\Phi^{-}\rangle_{12}$)} & \tabincell{l}{ $|\Phi^{-}\rangle_{34}$ ($|\Phi^{-}\rangle_{34}$)} & \tabincell{l}{ $|\Phi^{-}\rangle_{65}$ ($|\Phi^{+}\rangle_{65}$)}\\
        Corresponding keys &  $00$ ($01$) & $01$ ($01$) & $01$ ($00$)\\ \\

        \tabincell{l}{Possible measurement\\ outcomes} & \tabincell{l}{ $|\Phi^{+}\rangle_{12}$ ($|\Phi^{-}\rangle_{12}$)} & \tabincell{l}{ $|\Psi^{+}\rangle_{34}$ ($|\Psi^{+}\rangle_{34}$)} & \tabincell{l}{ $|\Psi^{+}\rangle_{65}$ ($|\Psi^{-}\rangle_{65}$)}\\
        Corresponding keys &  $00$ ($01$) & $10$ ($10$) & $10$ ($11$)\\ \\

        \tabincell{l}{Possible measurement\\ outcomes} & \tabincell{l}{ $|\Phi^{+}\rangle_{12}$ ($|\Phi^{-}\rangle_{12}$)} & \tabincell{l}{ $|\Psi^{-}\rangle_{34}$ ($|\Psi^{-}\rangle_{34}$)} & \tabincell{l}{ $|\Psi^{-}\rangle_{65}$ ($|\Psi^{+}\rangle_{65}$)}\\
        Corresponding keys &  $00$ ($01$) & $11$ ($11$) & $11$ ($10$)\\ \\

        \tabincell{l}{Possible measurement\\ outcomes} & \tabincell{l}{ $|\Psi^{+}\rangle_{12}$ ($|\Psi^{-}\rangle_{12}$)} & \tabincell{l}{ $|\Phi^{+}\rangle_{34}$ ($|\Phi^{+}\rangle_{34}$)} & \tabincell{l}{ $|\Psi^{+}\rangle_{65}$ ($|\Psi^{-}\rangle_{65}$)}\\
        Corresponding keys &  $10$ ($11$) & $00$ ($00$) & $10$ ($11$)\\ \\

        \tabincell{l}{Possible measurement\\ outcomes} & \tabincell{l}{ $|\Psi^{+}\rangle_{12}$ ($|\Psi^{-}\rangle_{12}$)} & \tabincell{l}{ $|\Phi^{-}\rangle_{34}$ ($|\Phi^{-}\rangle_{34}$)} & \tabincell{l}{ $|\Psi^{-}\rangle_{65}$ ($|\Psi^{+}\rangle_{65}$)}\\
        Corresponding keys &  $10$ ($11$) & $01$ ($01$) & $11$ ($10$)\\ \\

        \tabincell{l}{Possible measurement\\ outcomes} & \tabincell{l}{ $|\Psi^{+}\rangle_{12}$ ($|\Psi^{-}\rangle_{12}$)} & \tabincell{l}{ $|\Psi^{+}\rangle_{34}$ ($|\Psi^{+}\rangle_{34}$)} & \tabincell{l}{ $|\Phi^{+}\rangle_{65}$ ($|\Phi^{-}\rangle_{65}$)}\\
        Corresponding keys &  $10$ ($11$) & $10$ ($10$) & $00$ ($01$)\\ \\

        \tabincell{l}{Possible measurement\\ outcomes} & \tabincell{l}{ $|\Psi^{+}\rangle_{12}$ ($|\Psi^{-}\rangle_{12}$)} & \tabincell{l}{ $|\Psi^{-}\rangle_{34}$ ($|\Psi^{-}\rangle_{34}$)} & \tabincell{l}{ $|\Phi^{-}\rangle_{65}$ ($|\Phi^{+}\rangle_{65}$)}\\
        Corresponding keys &  $10$ ($11$) & $11$ ($11$) & $01$ ($00$)\\
        \noalign{\smallskip}\hline
        \end{tabular}}
\end{table}
        In order to do that, he should find out $P_{1}$'s key $K_{1}$ and $P_{3}$'s key $K_{3}$. However, he is unable to complete this task. From the Eq. (\ref{beUsed}) and the encoding arrangement Eq. (\ref{bEncoding}), he can just have $K_{1}\oplus K_{2}\oplus K_{3}=0$. $P_{2}$ is  unable to exactly figure out the values of $K_{1}$ and $K_{3}$. In the step (S\ref{step:last}), $P_{2}$ may get $C_{13}$ ($C_{12}$) that's the ciphertext of $M_{1}\oplus M_{3}$ ($M_{1}\oplus M_{2}$) encrypted by $K_{2}$ ($K_{3}$). But he cannot obtain $C_{13}$ and $C_{12}$ at the same time in our protocol. If $P_{2}$ gets $C_{12}$, he could compute $C_{12}\oplus C_{2}=C_{1}=K_{1}\oplus M_{1}$. He could also get $C_{3}$ and $C_{13}$, computing $C_{3}\oplus C_{13}=K_{3}\oplus M_{3} \oplus K_{1} \oplus M_{1} \oplus K_{3} \oplus M_{3}=K_{1}\oplus M_{1}$ which is the ciphertext of $M_{1}$ encrypted by $K_{1}$. Obviously, he cannot have the $P_{1}$'s private input $M_{1}$ because he does not know the exact value of $K_{1}$. $P_{2}$ also fails to get $P_{3}$'s private input $M_{3}$ because he cannot get $P_{3}$'s key $K_{3}$. Therefore, $P_{2}$ couldn't have the private inputs of $P_{1}$ and $P_{3}$. We take the two bits keys as an example so as to see the relations of the participants' keys and their measurement outcomes. The details see Table \ref{tab:1}. From this table, we know that $P_{2}$ cannot definitely determine the keys of $P_{1}$ and $P_{3}$ depending on his own key.\par

    Second, we will show that our protocol is still secure if $P_{1}$ wants to steal others' inputs. In the proposed protocol, the participant $P_{1}$ who prepares the state $\Psi_{6qb}$ is more powerful  than $P_{2}$ and $P_{3}$. He may prepare some particular fake particles, sending them to $P_{2}$ and $P_{3}$, respectively. And then he can determine the keys of $P_{2}$ and $P_{3}$ with certainty. Hence he could get the private inputs of $P_{2}$ and $P_{3}$. For instance, $P_{1}$ can send particles in the ordered sequence in Bell state $|\Psi^{-}\rangle$ ($|\Psi^{+}\rangle$) to $P_{2}$ ($P_{3}$) if he wishes to decide some two bits of $P_{2}$'s ($P_{3}$'s) key to be 11 (10). He could also prepare the real state $\Psi_{6qb}$, but then sends the particular particles to $P_{2}$ ($P_{3}$) according to Eqs.(\ref{twoSplit1}-\ref{twoSplit5}). He can finally steal the key of $P_{2}$ ($P_{3}$) depending on the relation of their measurement outcomes, obtaining their private inputs. Unfortunately, these attacks will be detected in the step (S\ref{step:colCheck}) of our protocol if $\delta$ is large enough. The general attack of $P_{1}$ can be described by a unitary operation: $U_{A}$ performed on qubits, including the state $\Psi_{6qb}$ and the probe state initialized as $|0\rangle_{A}$ before $P_{1}$ sends the particles to $P_{2}$ and $P_{3}$. We can prove that the final state of $\Psi_{6qb}$ would not be entangled with $P_{1}$'s probe state, which implies $P_{1}$ cannot get any information about exact measurement outcomes of $P_{2}$ and $P_{3}$ through his probe if there is no error to occur. Thus he could not get the keys of $P_{2}$ and $P_{3}$.\par
    The most general operation $P_{1}$ can do is to entangle the state $\Psi_{6qb}$ with the probe state initialized as $|0\rangle_{A}$, which can be written as
    \begin{eqnarray}\label{smit}
        U_{A}\Psi_{6qb}|0\rangle_{A}
        =\frac{1}{4}[
            |\Phi^{+}\rangle_{12}(&|\Phi^{+}\rangle_{34}&|\Phi^{+}\rangle_{65}|A_{1}\rangle
                                +|\Phi^{-}\rangle_{34}|\Phi^{-}\rangle_{65}|A_{2}\rangle \nonumber\\
                                +&|\Psi^{+}\rangle_{34}&|\Psi^{+}\rangle_{65}|A_{3}\rangle
                                +|\Psi^{-}\rangle_{34}|\Psi^{-}\rangle_{65}|A_{4}\rangle) \nonumber\\
            +|\Phi^{-}\rangle_{12}(-&|\Phi^{+}\rangle_{34}&|\Phi^{-}\rangle_{65}|A_{5}\rangle|
                                +|\Phi^{-}\rangle_{34}|\Phi^{+}\rangle_{65}|A_{6}\rangle \nonumber\\
                                -&|\Psi^{+}\rangle_{34}&|\Psi^{-}\rangle_{65}|A_{7}\rangle
                                +|\Psi^{-}\rangle_{34}|\Psi^{+}\rangle_{65}|A_{8}\rangle) \nonumber\\
            +|\Psi^{+}\rangle_{12}(&|\Phi^{+}\rangle_{34}&|\Psi^{+}\rangle_{65}|A_{9}\rangle
                                +|\Phi^{-}\rangle_{34}|\Psi^{-}\rangle_{65}|A_{10}\rangle \nonumber\\
                                -&|\Psi^{+}\rangle_{34}&|\Phi^{+}\rangle_{65}|A_{11}\rangle
                                -|\Psi^{-}\rangle_{34}|\Phi^{-}\rangle_{65}|A_{12}\rangle) \nonumber\\
            +|\Psi^{-}\rangle_{12}(-&|\Phi^{+}\rangle_{34}&|\Psi^{-}\rangle_{65}|A_{13}\rangle
                                +|\Phi^{-}\rangle_{34}|\Psi^{+}\rangle_{65}|A_{14}\rangle \nonumber\\
                                +&|\Psi^{+}\rangle_{34}&|\Phi^{-}\rangle_{65}|A_{15}\rangle
                                -|\Psi^{-}\rangle_{34}|\Phi^{+}\rangle_{65}|A_{16}\rangle)],
    \end{eqnarray}
    where $|A_{i}\rangle$ ($i=1, 2,\ldots, 16$) are some unnormalized states in $P_{1}$'s probe space. We will prove that that the final state of $\Psi_{6qb}$ would not be entangled with $P_{1}$'s probe state if he can escape the detection in the step (S\ref{step:colCheck}) of our protocol. \par
    On one hand, if $P_{1}$ measures the particles (1, 2) with the basis $\{|0\rangle, |1\rangle\}$, then $P_{2}$ measures the particle 3 with the basis $\{|0\rangle, |1\rangle\}$, and $P_{2}$ and $P_{3}$ measure the particles (4, 5, 6) with the basis $\{|\gamma_{j}^{1}\rangle|j=1,2,\ldots,8\}$. Then, the state in Eq. (\ref{smit}) can be rewritten as follows:

    \begin{eqnarray}\label{01check}
    \frac{1}{\sqrt{8}}\{|000\rangle[|\gamma_{1}^{1}\rangle(|A_{1}\rangle&+&|A_{3}\rangle+|A_{6}\rangle+|A_{8}\rangle)
    +|\gamma_{4}^{1}\rangle(|A_{2}\rangle-|A_{4}\rangle-|A_{5}\rangle+|A_{7}\rangle)\nonumber\\
    +|\gamma_{6}^{1}\rangle(|A_{1}\rangle&-&|A_{3}\rangle+|A_{6}\rangle-|A_{8}\rangle)
    +|\gamma_{7}^{1}\rangle(|A_{2}\rangle+|A_{4}\rangle-|A_{5}\rangle-|A_{7}\rangle)] \nonumber\\
    +|001\rangle[|\gamma_{2}^{1}\rangle(|A_{2}\rangle&+&|A_{4}\rangle+|A_{5}\rangle+|A_{7}\rangle)
    +|\gamma_{3}^{1}\rangle(|A_{1}\rangle+|A_{3}\rangle-|A_{6}\rangle-|A_{8}\rangle)\nonumber\\
    +|\gamma_{5}^{1}\rangle(|A_{4}\rangle&-&|A_{2}\rangle-|A_{5}\rangle+|A_{7}\rangle)
    +|\gamma_{8}^{1}\rangle(|A_{1}\rangle+|A_{3}\rangle-|A_{6}\rangle-|A_{8}\rangle)] \nonumber\\
    +|010\rangle[|\gamma_{3}^{1}\rangle(|A_{10}\rangle&+&|A_{12}\rangle+|A_{13}\rangle+|A_{15}\rangle)
    +|\gamma_{2}^{1}\rangle(|A_{9}\rangle-|A_{11}\rangle-|A_{14}\rangle+|A_{16}\rangle)\nonumber\\
    +|\gamma_{5}^{1}\rangle(|A_{9}\rangle&+&|A_{11}\rangle-|A_{14}\rangle-|A_{16}\rangle)
    +|\gamma_{8}^{1}\rangle(|A_{10}\rangle-|A_{12}\rangle+|A_{13}\rangle-|A_{15}\rangle)] \nonumber\\
    +|011\rangle[|\gamma_{4}^{1}\rangle(|A_{9}\rangle&+&|A_{11}\rangle+|A_{14}\rangle+|A_{16}\rangle)
    +|\gamma_{1}^{1}\rangle(|A_{12}\rangle-|A_{10}\rangle+|A_{13}\rangle-|A_{15}\rangle)\nonumber\\
    +|\gamma_{6}^{1}\rangle(|A_{10}\rangle&+&|A_{12}\rangle-|A_{13}\rangle-|A_{15}\rangle)
    +|\gamma_{7}^{1}\rangle(|A_{11}\rangle-|A_{9}\rangle-|A_{14}\rangle+|A_{16}\rangle)] \nonumber\\
    -|100\rangle[|\gamma_{5}^{1}\rangle(|A_{9}\rangle&+&|A_{11}\rangle+|A_{14}\rangle+|A_{16}\rangle)
    +|\gamma_{2}^{1}\rangle(|A_{9}\rangle-|A_{11}\rangle+|A_{14}\rangle-|A_{16}\rangle)\nonumber\\
    +|\gamma_{3}^{1}\rangle(|A_{10}\rangle&+&|A_{12}\rangle-|A_{13}\rangle-|A_{15}\rangle)
    +|\gamma_{8}^{1}\rangle(|A_{10}\rangle-|A_{12}\rangle-|A_{13}\rangle+|A_{15}\rangle)] \nonumber\\
    -|101\rangle[|\gamma_{6}^{1}\rangle(|A_{10}\rangle&+&|A_{12}\rangle+|A_{13}\rangle+|A_{15}\rangle)
    +|\gamma_{1}^{1}\rangle(|A_{12}\rangle-|A_{10}\rangle-|A_{13}\rangle+|A_{15}\rangle)\nonumber\\
    +|\gamma_{4}^{1}\rangle(|A_{9}\rangle&+&|A_{11}\rangle-|A_{14}\rangle-|A_{16}\rangle)
    +|\gamma_{7}^{1}\rangle(|A_{11}\rangle-|A_{9}\rangle+|A_{14}\rangle-|A_{16}\rangle)] \nonumber\\
    +|110\rangle[|\gamma_{7}^{1}\rangle(|A_{2}\rangle&+&|A_{4}\rangle+|A_{5}\rangle+|A_{7}\rangle)
    +|\gamma_{1}^{1}\rangle(|A_{1}\rangle+|A_{3}\rangle-|A_{6}\rangle-|A_{8}\rangle)\nonumber\\
    +|\gamma_{4}^{1}\rangle(|A_{2}\rangle&-&|A_{4}\rangle+|A_{5}\rangle-|A_{7}\rangle)
    +|\gamma_{6}^{1}\rangle(|A_{1}\rangle-|A_{3}\rangle-|A_{6}\rangle+|A_{8}\rangle)] \nonumber\\
    +|111\rangle[|\gamma_{8}^{1}\rangle(|A_{1}\rangle&+&|A_{3}\rangle+|A_{6}\rangle+|A_{8}\rangle)
    +|\gamma_{2}^{1}\rangle(|A_{2}\rangle+|A_{4}\rangle-|A_{5}\rangle-|A_{7}\rangle)\nonumber\\
    +|\gamma_{3}^{1}\rangle(|A_{3}\rangle&-&|A_{1}\rangle-|A_{6}\rangle+|A_{8}\rangle)
    +|\gamma_{5}^{1}\rangle(|A_{4}\rangle-|A_{2}\rangle+|A_{5}\rangle-|A_{7}\rangle)]\}_{123456A}. \nonumber\\
    \end{eqnarray}
    According to Eq. (\ref{first}), if $P_{1}$ introduces no error, the following conditions should be satisfied:
    \begin{eqnarray}\label{01should}
      |A_{1}\rangle&=&|A_{3}\rangle=|A_{6}\rangle=|A_{8}\rangle, \nonumber\\
      |A_{2}\rangle&=&|A_{4}\rangle=|A_{5}\rangle=|A_{7}\rangle, \nonumber\\
      |A_{9}\rangle&=&|A_{11}\rangle=|A_{14}\rangle=|A_{16}\rangle, \nonumber\\
      |A_{10}\rangle&=&|A_{12}\rangle=|A_{13}\rangle=|A_{5}\rangle.
    \end{eqnarray}
    On the other hand, if $P_{1}$ measures the particles (1, 2) with the basis $\{|+\rangle, |-\rangle\}$, then $P_{2}$ measures the particle 3 with the basis $\{|+\rangle, |-\rangle\}$, and $P_{2}$ and $P_{3}$ measure the particles (4, 5, 6) with the basis $\{|\gamma_{j}^{2}\rangle|j=1,2,\ldots,8\}$. Then, the state in Eq. (\ref{smit}) can be rewritten as follows:
    \begin{eqnarray}\label{+-check}
    \frac{1}{\sqrt{8}}\{|+++\rangle[|\gamma_{1}^{2}\rangle(|A_{2}\rangle&+&|A_{3}\rangle+|A_{13}\rangle+|A_{16}\rangle)
    +|\gamma_{4}^{2}\rangle(|A_{1}\rangle-|A_{4}\rangle+|A_{14}\rangle-|A_{15}\rangle)\nonumber\\
    +|\gamma_{6}^{2}\rangle(|A_{3}\rangle&-&|A_{2}\rangle+|A_{13}\rangle-|A_{16}\rangle)
    +|\gamma_{7}^{2}\rangle(|A_{1}\rangle+|A_{4}\rangle-|A_{14}\rangle-|A_{15}\rangle)] \nonumber\\
    +|++-\rangle[|\gamma_{2}^{2}\rangle(|A_{1}\rangle&+&|A_{4}\rangle+|A_{14}\rangle+|A_{15}\rangle)
    +|\gamma_{3}^{2}\rangle(|A_{3}\rangle-|A_{2}\rangle-|A_{13}\rangle+|A_{16}\rangle)\nonumber\\
    +|\gamma_{5}^{2}\rangle(|A_{1}\rangle&-&|A_{4}\rangle-|A_{14}\rangle+|A_{15}\rangle)
    +|\gamma_{8}^{2}\rangle(|A_{2}\rangle+|A_{3}\rangle-|A_{13}\rangle-|A_{15}\rangle)] \nonumber\\
    +|+-+\rangle[|\gamma_{3}^{2}\rangle(|A_{5}\rangle&+&|A_{8}\rangle+|A_{10}\rangle+|A_{11}\rangle)
    +|\gamma_{2}^{2}\rangle(|A_{6}\rangle-|A_{7}\rangle+|A_{9}\rangle-|A_{12}\rangle)\nonumber\\
    +|\gamma_{5}^{2}\rangle(|A_{6}\rangle&+&|A_{7}\rangle-|A_{9}\rangle-|A_{12}\rangle)
    +|\gamma_{8}^{2}\rangle(|A_{8}\rangle-|A_{5}\rangle+|A_{10}\rangle-|A_{11}\rangle)] \nonumber\\
    +|+--\rangle[|\gamma_{4}^{2}\rangle(|A_{6}\rangle&+&|A_{7}\rangle+|A_{9}\rangle+|A_{12}\rangle)
    +|\gamma_{1}^{2}\rangle(|A_{8}\rangle-|A_{5}\rangle-|A_{10}\rangle+|A_{11}\rangle)\nonumber\\
    +|\gamma_{6}^{2}\rangle(|A_{5}\rangle&+&|A_{8}\rangle-|A_{10}\rangle-|A_{11}\rangle)
    +|\gamma_{7}^{2}\rangle(|A_{6}\rangle-|A_{7}\rangle-|A_{9}\rangle+|A_{12}\rangle)] \nonumber\\
    +|-++\rangle[|\gamma_{5}^{2}\rangle(|A_{6}\rangle&+&|A_{7}\rangle+|A_{9}\rangle+|A_{12}\rangle)
    +|\gamma_{2}^{2}\rangle(|A_{6}\rangle-|A_{7}\rangle-|A_{9}\rangle+|A_{12}\rangle)\nonumber\\
    +|\gamma_{3}^{2}\rangle(|A_{5}\rangle&+&|A_{8}\rangle-|A_{10}\rangle-|A_{11}\rangle)
    +|\gamma_{8}^{2}\rangle(|A_{8}\rangle-|A_{5}\rangle-|A_{10}\rangle+|A_{11}\rangle)] \nonumber\\
    +|-+-\rangle[|\gamma_{6}^{2}\rangle(|A_{5}\rangle&+&|A_{8}\rangle+|A_{10}\rangle+|A_{11}\rangle)
    +|\gamma_{1}^{2}\rangle(|A_{8}\rangle-|A_{5}\rangle+|A_{10}\rangle-|A_{11}\rangle)\nonumber\\
    +|\gamma_{4}^{2}\rangle(|A_{6}\rangle&+&|A_{7}\rangle-|A_{9}\rangle-|A_{12}\rangle)
    +|\gamma_{7}^{2}\rangle(|A_{6}\rangle-|A_{7}\rangle+|A_{9}\rangle-|A_{12}\rangle)] \nonumber\\
    +|--+\rangle[|\gamma_{7}^{2}\rangle(|A_{1}\rangle&+&|A_{4}\rangle+|A_{14}\rangle+|A_{15}\rangle)
    +|\gamma_{1}^{2}\rangle(|A_{2}\rangle+|A_{3}\rangle-|A_{13}\rangle-|A_{16}\rangle)\nonumber\\
    +|\gamma_{4}^{2}\rangle(|A_{1}\rangle&-&|A_{4}\rangle-|A_{14}\rangle+|A_{15}\rangle)
    +|\gamma_{6}^{2}\rangle(|A_{3}\rangle-|A_{2}\rangle+|A_{13}\rangle-|A_{16}\rangle)] \nonumber\\
    +|---\rangle[|\gamma_{8}^{2}\rangle(|A_{2}\rangle&+&|A_{3}\rangle+|A_{13}\rangle+|A_{16}\rangle)
    +|\gamma_{2}^{2}\rangle(|A_{1}\rangle+|A_{4}\rangle-|A_{14}\rangle-|A_{15}\rangle)\nonumber\\
    +|\gamma_{3}^{2}\rangle(|A_{3}\rangle&-&|A_{2}\rangle+|A_{13}\rangle-|A_{16}\rangle)
    +|\gamma_{5}^{2}\rangle(|A_{1}\rangle-|A_{4}\rangle+|A_{14}\rangle-|A_{15}\rangle)]\}_{123456A}. \nonumber\\
    \end{eqnarray}
    According to Eq. (\ref{second}), if $P_{1}$ introduces no error, the following conditions should be satisfied:
    \begin{eqnarray}\label{+-should}
      |A_{1}\rangle&=&|A_{4}\rangle=|A_{14}\rangle=|A_{15}\rangle, \nonumber\\
      |A_{2}\rangle&=&|A_{3}\rangle=|A_{13}\rangle=|A_{16}\rangle, \nonumber\\
      |A_{5}\rangle&=&|A_{8}\rangle=|A_{10}\rangle=|A_{11}\rangle, \nonumber\\
      |A_{6}\rangle&=&|A_{7}\rangle=|A_{9}\rangle=|A_{12}\rangle.
    \end{eqnarray}
    We can derive from Eq.~(\ref{01should}) and Eq.~(\ref{+-should}) that $|A_{1}\rangle = |A_{2}\rangle = \ldots = |A_{16}\rangle$, which means that the state $\Psi_{6qb}$ and the prob state prepared by $P_{1}$ are entirely not entangled. Thus the subsequent measurement outcome of the prob state tells $P_{1}$ nothing. \par
     $P_{1}$ may use the similar attack as $P_{2}$ to derive the inputs of $P_{2}$ and $P_{3}$ in accordance with the cipertexts $C_{2}$ and $C_{3}$ that are $M_{2}$ and $M_{3}$ encoded with $K_{2}$ and $K_{3}$, respectively. However, according to the Eq. (\ref{beUsed}) and the encoding arrangement Eq. (\ref{bEncoding}), he can have $K_{1}\oplus K_{2}\oplus K_{3}=0$ but the keys $K_{2}$ and $K_{3}$ and therefore is unable to steal $M_{2}$ and $M_{3}$ offered by $P_{2}$ and $P_{3}$, respectively. So the protocol remains secure against this attack.\par
    In the step (S\ref{step:last}) of our protocol, $P_{1}$ first determines whether $M_{2}=M_{3}$ using his own key $K_{1}$, $P_{2}$ ($P_{3}$) then determines if $M_{1}=M_{3}$ ($M_{1}=M_{2}$) or not based on his key $K_{2}$ ($K_{3}$) and at last they can get the comparison result. In fact, the order in which one participant decides whether or not the other two participants' private inputs are equal is not important because their private inputs are encrypted by their keys that are kept secret.\par
    Unfortunately, any two participants can collude with each other to derive the third one's key according to $K_{1} \oplus K_{2} \oplus K_{3}=0$, obtaining the corresponding private input. So it would be interesting to design multi-party quantum private comparison protocols that are still secure against such an attack.\par
    \subsubsection{Security analysis over lossy and noise channel}
    In the above analysis, the quantum channels are assumed to be under the ideal condition (i.e. noiseless and lossless). But quantum channels are usually lossy and noisy in the real world. In this section, we show that our protocol remains secure in lossy and noisy quantum channels. The eavesdropper, Eve, is assumed to be powerful enough to establish  an ideal channel with any participant. We discuss the lossy and noisy quantum channels in case as follows.

    \textbf{Case 1. Lossy quantum channel}

    In such a quantum channel, Eve may intercepts particles sent from $P_{1}$ to $P_{2}$ and $P_{3}$. She then keeps some of them and transmits the other particles to $P_{2}$ and $P_{3}$ through an ideal channel. If the intercepted particles are not decoy particles, she is able to perform measurements on the related particles with Bell basis. The measurement results will lead to the leakage of the some key bits for $P_{1}$, she will finally get some information of $P_{1}$'s private input. Fortunately, our protocol remains secure against such a attack. In the Step 2 of our protocol, $P_{2}$ ($P_{3}$) informs $P_{1}$ which particles have been received and which are lost during the transmission. $P_{1}$ and $P_{2}$ ($P_{3}$) only employ the received particles to make a public discussion and finish equality comparison. The intercepted particles are useless and Eve will fail to extract any information about $P_{1}$'s key that is used to encrypt his private input.

    \textbf{Case 2. Noisy quantum channel}

    Eve can intercept the particles sent from $P_{1}$ to $P_{2}$ ($P_{3}$), performing intercept-resend attack or entangle-measure attack, forwarding these tampered particles to  $P_{2}$ ($P_{3}$) through an idea channel established by herself. In this situation, Eve tries her best to cover up the tampering of particles as the noise existed on the quantum channel between  $P_{1}$ and $P_{2}$ ($P_{3}$). We have learned that these attacks will be caught if the eavesdropper detection rate of our protocol is smaller than the quantum bit error rate of noise (QBER). In accordance with \cite{jennewein2000quantum,hughes2002practical,gobby2004quantum}, the QBER is roughly between $2\%$ and $8.9\%$ depending on the different channel situations (e.g., distance, etc.). Fortunately, the detection rate for decoy particle in our protocol is $25\%$ that is greater than the error rate of the quantum channel. Therefore, our protocol is also secure in the noisy quantum channel.

    Up until now, we have completed the analysis of the correctness and the security of our protocol.

    Note that if Lin et al.'s protocol \cite{lin2014quantum} is used for three participants' private comparison, it needs run twice in the worst case. But our protocol needs run only once even in the worst case. For Chang et al.'s protocol \cite{chang2013multi}, in the case of the number of the participants is 3, our protocol needs more $6\delta$ particles than their protocol do and our protocol will suffer collusion attack as mentioned in the previous security analysis. This is the disadvantages of our protocol! However, such disadvantages happen because of the lack of an external third party. The more $6\delta$ particles serves as participant attack detection, because the initial quantum states are prepared by $P_{1}$. We need to check if he is honest. If we include the external third party who is assumed to be semi-honest and prepares the initial quantum state, these defects will vanish. But it will require more various quantum states and different quantum operations. Chang et al.'s protocol may not be able to finish the private task if it lacks the semi-honest third party. In practice, resorting to an external third party for help in quantum private comparison may unexpectedly result in some private information leakage and this help is usually not free. In this sense, designing quantum private comparison protocols without an external third party is necessary.

\section{Conclusions}
    We present a new three-party quantum private comparison protocol based on genuinely maximally entangled six-qubit states. Three participants can determine whether their private information are equal without the assistance of an external third party and in the meantime keep their inputs secret, respectively. The proposed protocol does not require any unitary operations to encode information for the sake of the excellent properties of genuinely maximally entangled six-qubit states. Because the proposed protocol utilizes one-step quantum transmission, it can be prevented from Trojan horse attacks. Finally, we also analyze the correctness and security of our protocol.

\begin{acknowledgements}
This work is supported by the National Natural Science Foundation of China under Grants No.61272013.
\end{acknowledgements}


\bibliographystyle{spphys}       
\bibliography{QuantumEndnoteForQPC}

\end{document}